# Compound Heat Wave and PM$_{2.5}$ Pollution Episodes in U.S. Cities

SARAH HENRY*

*National Weather Center Research Experiences for Undergraduates Program*
*Norman, Oklahoma;*
*University of Illinois at Urbana-Champaign*
*Champaign, Illinois*

CHENGHAO WANG

*School of Meteorology, University of Oklahoma*
*Norman, Oklahoma;*
*Department of Geography and Environmental Sustainability, University of Oklahoma*
*Norman, Oklahoma*

ABSTRACT

This study analyzes heat waves (HWs), air pollution (AP) episodes, and compound HW and AP events (CE) in the urban environment and provides a comparison between events in urban areas (UAs) and rural areas (RAs). A 1-km gridded daily minimum temperature dataset and a 1-km gridded daily PM$_{2.5}$ concentration dataset were used along with geospatial data to characterize events by their frequency, intensity in heat, intensity in pollution, and duration. The greatest differences between UAs and RAs in frequency, heat intensity, pollution intensity, and duration for all events were seen in the West and Southwest regions. For both UAs and RAs, it was found that HWs were the most frequent, intense, and longest lasting in the West and Southwest regions, AP episodes were the most frequent and longest lasting in the Northeast, Ohio Valley, and Southeast regions, and AP episodes were the most intense in the Northern Rockies and Plains and Upper Midwest regions. It was concluded that HWs (AP episodes) had a greater impact on CEs than AP episodes (HWs) in regions with more prominent HWs (AP episodes).

## 1. Introduction

Heat waves (HWs) and PM$_{2.5}$ air pollution (AP) episodes are deadly events that increase rates of mortality and hospital admissions (Patz et al. 2005; Anderson et al. 2012). HWs are one of the deadliest weather-related disasters, with more than 1,300 heat-related deaths each year in the United States (U.S. EPA OAR 2016). With rising global temperatures, it is becoming increasingly important to understand extreme high-temperature events and how they relate to other extreme events, such as AP episodes.

HWs are caused by atmospheric blocking events, where air becomes stagnant and hotter from increased solar radiation (Perkins 2015). The extreme heat can be long lasting and deadly, with significant examples being Chicago in 1995, Europe in 2003, Russia in 2010, and Pakistan in 2015 (Browning et al. 2006; United Nations Environment Programme 2003; Dole et al. 2011; Masood et al. 2015). HWs also have disastrous impacts on agricultural production (Lobell et al. 2011; Coffel et al. 2019), can compound with drought to cause wildfires (Zscheischler et al. 2020), and increase energy demand for cooling (Zuo et al. 2015). Due to global warming, HWs are projected to be more frequent, more intense, and longer lasting in the 21st century (Meehl and Tebaldi 2004). A previous study (Alexander et al. 2006) found that increases in daily minimum temperatures have outpaced increases in daily maximum temperatures from 1951 to 2003. Therefore, a deeper understanding of the drivers and impacts of HW events is required to mitigate these changes. Within the urban environment, the impacts of HWs can be amplified by the urban heat island (UHI) effect. The UHI effect is a phenomenon where surface temperature and/or near-surface temperature are elevated in urban areas (UAs) relative to surrounding rural areas (RAs) (Phelan et al. 2015), which is primarily caused by unique thermal properties of construction materials, lack of vegetation, building geometry, and increased anthropogenic heat emissions. The interactions between HWs and the UHI have long been shown to increase heat-related mortality (Buechley et al. 1972; Tan et al. 2010), and recent studies have demonstrated their synergistic relationship, showing that their combined effects are greater

---
*Corresponding author address:* Sarah Henry, University of Illinois at Urbana-Champaign, Champaign, IL 61801
E-mails: sarmariehenry@gmail.com (Henry); chenghao.wang@ou.edu (Wang)

than their sum (Tan et al. 2010; Li and Bou-Zeid 2013; Ramamurthy et al. 2017; Zhao et al. 2018). The impact of an HW amplified by UHI is typically greatest at night, when anthropogenic activities keep temperatures elevated when surrounding RAs cool down (Phelan et al. 2015). The background climate of a city affects the UHI, and while increasing trends of HW occurrence have been observed in some areas (Peterson et al. 2013; Smith et al. 2013), it is unknown if these patterns are true for all UAs.

AP episodes are days when levels of airborne pollutants are elevated. The most common and most adverse to public health pollutant is particulate matter less than 2.5 micrometers in diameter ($PM_{2.5}$) which has been linked to increasing risks of cardiovascular and respiratory diseases and death (Xing et al. 2016; Apte et al. 2015). Within UAs, $PM_{2.5}$ concentrations can be higher relative to surrounding RAs, creating an urban pollution island (UPI) similar to UHI (Cohen et al. 2005). In many areas, AP episodes are decreasing in frequency, intensity, and duration due to regional and/or national legislation increasing restrictions on pollutant emissions (World Health Organization 2021; U.S. EPA OAR 2022). However, wildfire events, which are expected to increase in frequency due to global warming (Tucker 2000), can potentially increase their concentrations.

The purpose of this study is to provide an understanding of how HWs and $PM_{2.5}$ pollution episodes vary in UAs across the contiguous U.S. (CONUS) and to explore compound HW and AP events (CE), where $PM_{2.5}$ concentrations are elevated during an HW. We aim to characterize the frequency, intensity, and duration of each type of event and identify UAs and climate regions most vulnerable to these events. Section 2 details data sources and methods used for this study. Section 3 presents the results of HW, AP episode, and compound HW and AP episode analysis in UAs in the US. Section 4 provides a discussion of the results found. Section 5 summarizes and concludes the findings of this study and discusses potential areas of future research.

## 2. Data and methods

This study analyzes 481 UAs, defined as a densely populated geographical area with a population size of 50,000 or more in the CONUS (U.S. Census Bureau 2010). The analysis period is from May to September from 2003 to 2016, which was constrained by data availability. HWs are identified using a daily minimum temperature ($T_{min}$) dataset from a 1-km global dataset of daily maximum and minimum near-surface air temperature (2003–2020) (Zhang et al. 2022). This temperature dataset provides near-surface air temperature with a spatial resolution of 1-km and is derived from satellite remote sensing products and ground-based weather station data. The 1-km grid is capable of capturing UHI effects. AP episodes are identified using a daily $PM_{2.5}$ concentration dataset (2000–2016) (Di et al. 2019). This dataset was developed using an ensemble model of machine learning algorithms and has a spatial resolution of 1-km over the CONUS. For each UA, a surrounding RA is defined, as well as a background area (BA) that surrounds both the UA and RA. RAs exclude other UAs, and BAs exclude other UAs and RAs. All areas are approximately equal in landmass and exclude water bodies.

HWs, AP episodes, and CEs are identified, then compared between UA and RA, and then compared between climate regions, as defined by the NOAA U.S. Climate Regions (NOAA 2023). For this study, an HW is defined as a period of three or more days where the daily $T_{min}$ exceeds the daily 90th percentile $T_{min}$ defined by the BA (Wang et al. 2020). The purpose of using the BA to determine the threshold is to compare UAs and RAs relative to their background climate. HW events are identified independently between UA and RA and then compared in their frequency, heat intensity, and duration. The frequency is defined as the number of events. The heat intensity of the HW is measured as the difference between the daily $T_{min}$ and the daily 90th percentile $T_{min}$ from the BA. The duration is defined as the number of days in the HW. An AP episode is defined as a period of one or more days where the daily $PM_{2.5}$ concentration exceeds 15 $\mu g/m^3$, which is the World Health Organization's 24-hour Air Quality Guideline level (World Health Organization 2021). AP episodes are identified independently between UA and RA and then compared in their frequency, pollution intensity, and duration. Frequency and duration are defined the same as for HWs. Pollution intensity is measured as the difference between the daily $PM_{2.5}$ concentration and 15 $\mu g/m^3$. A CE is defined as a period of one or more days within an HW where the daily $PM_{2.5}$ concentration exceeds 15 $\mu g/m^3$. CEs are identified independently between UA and RA and then compared in their frequency, heat intensity, pollution intensity, and duration, using the same definition as above.

## 3. Results

### a. Heat waves

The comparison of HW frequency, heat intensity, and duration between each UA and RA is shown in Figure 1. It was found that 99.17% of frequencies, 66.53% of heat intensities, and 95.84% of durations were greater in UA than RA.

The comparison of HW frequency, heat intensity, and duration of UAs and RAs between different regions is shown in Figure 2. On average, all region had greater HW frequency, intensity, and duration in UAs than in RAs except for the Upper Midwest region, where RAs had a greater average heat intensity than UAs. Specifically, 62.22% of RAs had a greater heat intensity than UAs in the Upper Midwest. The West and Southwest regions had the most frequent HWs on average for both UAs and RAs. The West and Upper Midwest regions had the most intense HWs on average for both UAs and RAs. The Southwest and West regions had the longest lasting HWs on average for both UAs and RAs.

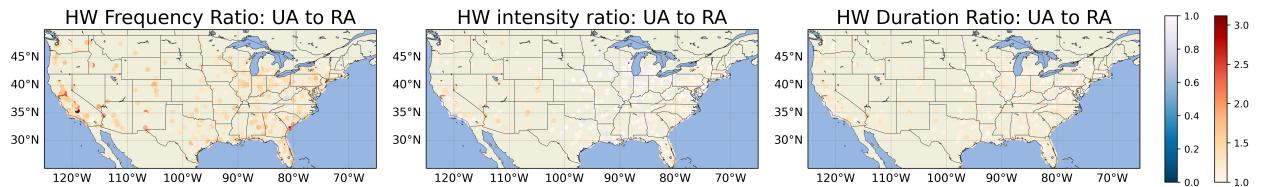

FIG. 1. Comparison between UA and RA HW frequency, heat intensity, and duration. A ratio of less than one indicates greater RA features, and a ratio of greater than one indicates greater UA features.

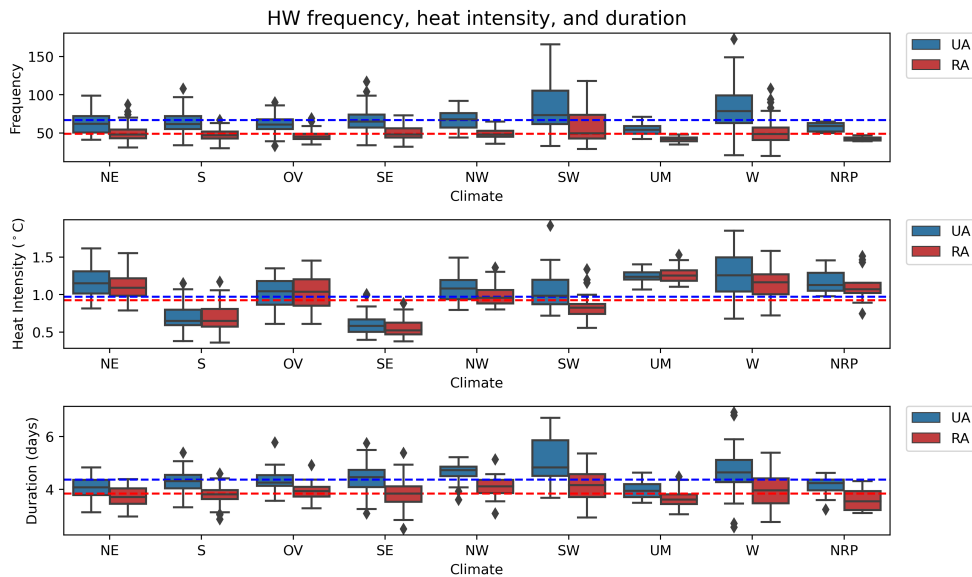

FIG. 2. Regional comparison of HWs frequency, heat intensity, and duration for UAs and RAs based on NOAA's 9 climate regions: Northeast (NE), South (S), Ohio Valley (OV), Southeast (SE), Northwest (NW), Southwest (SW), Upper Midwest (UM), West (W), Northern Rockies and Plains (NRP). Red dashed line indicates RA average, and blue dashed line indicates UA average. The center line of each box is the median, the box extends from lower to upper quartiles, vertical lines denote 1.5 times the interquartile range, and diamonds are outliers.

### b. $PM_{2.5}$ pollution episodes

The comparison of AP episode frequency, pollution intensity, and duration between each UA and RA is shown in Figure 3. It was found that 79.42% of frequencies, 59.46% of pollution intensities, and 71.52% of durations were greater in UA than RA.

The comparison of AP episode frequency, pollution intensity, and duration of UAs and RAs between different regions is shown in Figure 4. Most regions had similar frequencies, pollution intensities, and durations of AP events between UAs and RAs on average. The Northeast, Ohio Valley, and Southeast regions had the most frequent AP events on average for both UAs and RAs. The Northeast, Upper Midwest, and Northern Rockies and Plains regions had the greatest pollution intensity on average for both UAs and RAs. The Ohio Valley and Southeast regions had the longest lasting AP events on average for both UAs and RAs.

### c. Compound heat wave and air pollution episodes

The comparison of CE frequency, heat intensity, pollution intensity, and duration between each UA and RA is shown in Figure 5. It was found that 90.64% of frequencies, 97.71% of heat intensities, 77.34% of pollution intensities, and 56.96% of durations were greater in UA than RA.

The comparison of CE frequency, heat intensity, pollution intensity, and duration of UAs and RAs between different regions is shown in Figure 6. Each region had a higher average frequency and average heat intensity in UAs than in RAs, while each region had similar average pollution intensity and average duration between UAs and RAs. The Northeast and Ohio Valley regions had the most frequent CEs on average for both UAs and RAs. The West and Southwest regions had the greatest heat intensity on average for both UAs and RAs. The Northeast, Ohio Valley, and Southeast regions had the greatest pollution intensity on average, as well as the longest lasting CEs on average for both UAs and RAs.

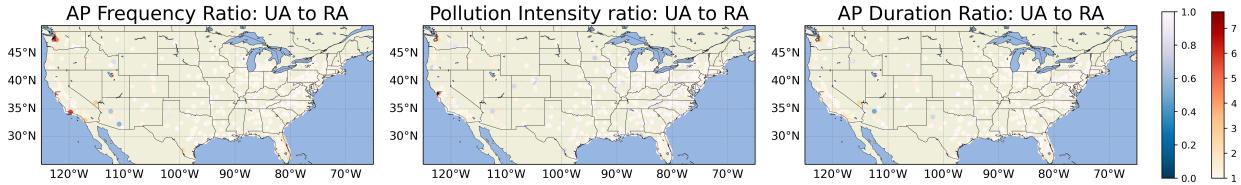

FIG. 3. Comparison between UA and RA AP episode frequency, pollution intensity, and duration. A ratio of less than one indicates greater RA features, and a ratio of greater than one indicates greater UA features.

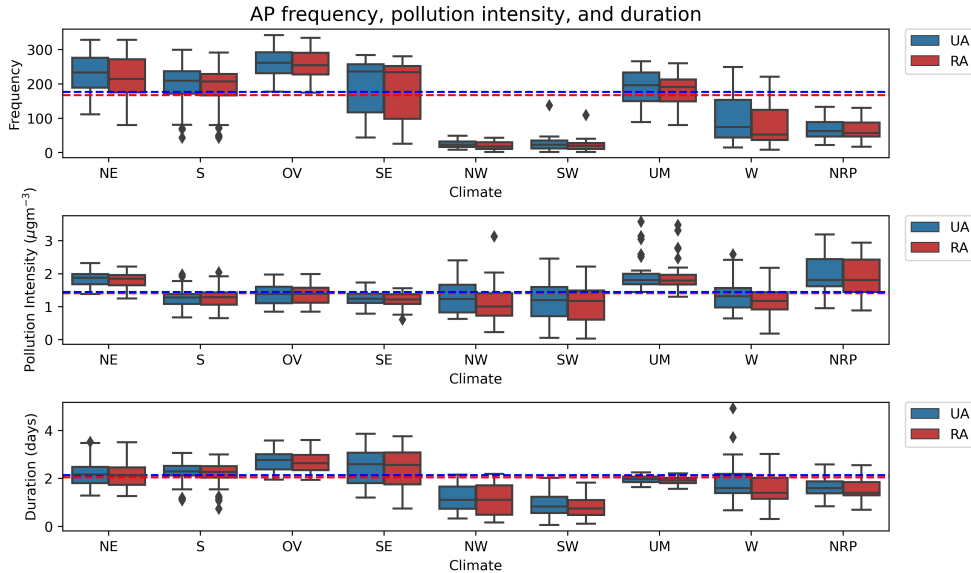

FIG. 4. Regional comparison of AP episode frequency, pollution intensity, and duration for UAs and RAs based on NOAA's 9 climate regions: Northeast (NE), South (S), Ohio Valley (OV), Southeast (SE), Northwest (NW), Southwest (SW), Upper Midwest (UM), West (W), Northern Rockies and Plains (NRP). Red dashed line indicates RA average, and blue dashed line indicates UA average. The center line of each box is the median, the box extends from lower to upper quartiles, vertical lines denote 1.5 times the interquartile range, and diamonds are outliers.

## 4. Discussion

### a. Heat waves

A greater frequency of HWs in 99.17% of UAs and a greater duration of HWs in 95.84% of UAs as compared to their rural counterparts, or RAs, is consistent with the synergistic relationship between the UHI and HWs (Tan et al. 2010; Li and Bou-Zeid 2013; Ramamurthy et al. 2017; Zhao et al. 2018). However, that relationship is not clear in the heat intensities of HWs. This could be an effect of the analysis performed, where all events were identified rather than concurrent events between both UA and RA.

For example, if an HW occurred in a UA but not in the surrounding RA, that identified HW could be less intense in heat and a result of the UHI. Future research could analyze only HWs that occurred in both UAs and RAs. Additionally, the definition of HW used impacted the results. $T_{min}$ was used, as HWs have their greatest impact at night when there is little or no relief from elevated temperatures (Phelan et al. 2015). This identifies HWs that would impact public health but may not identify the greatest intensity of heat during an HW. Future research could analyze HWs with a definition based on $T_{max}$ and/or other temperature thresholds, such as a percentile greater than the 90th percentile, or a fixed temperature threshold.

It was found that HW frequencies, heat intensities, and durations were greatest in the West and Southwest regions, which is consistent with other research and has been demonstrated to be driven by anthropogenic activities (Lopez et al. 2018). Furthermore, it has been projected that the intensity of HWs will increase in the West and South throughout the 21st century (Meehl and Tebaldi 2004). It was found that the heat intensity was greater in 62.22% of RAs than UAs in the Upper Midwest region. A recent study found that the Midwest experiences more intense HWs than other regions (Lopez et al. 2018), and this pattern could be true for both UAs and RAs. Further research could explore local climate regions and analyze how background climate impacts HWs in both UAs and RAs.

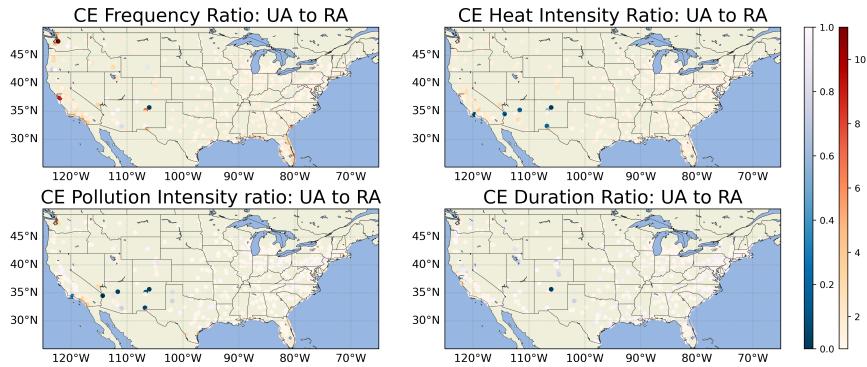

FIG. 5. Comparison between UA and RA CE frequency, heat intensity, pollution intensity, and duration. A ratio of less than one indicates greater RA features, and a ratio of greater than one indicates greater UA features.

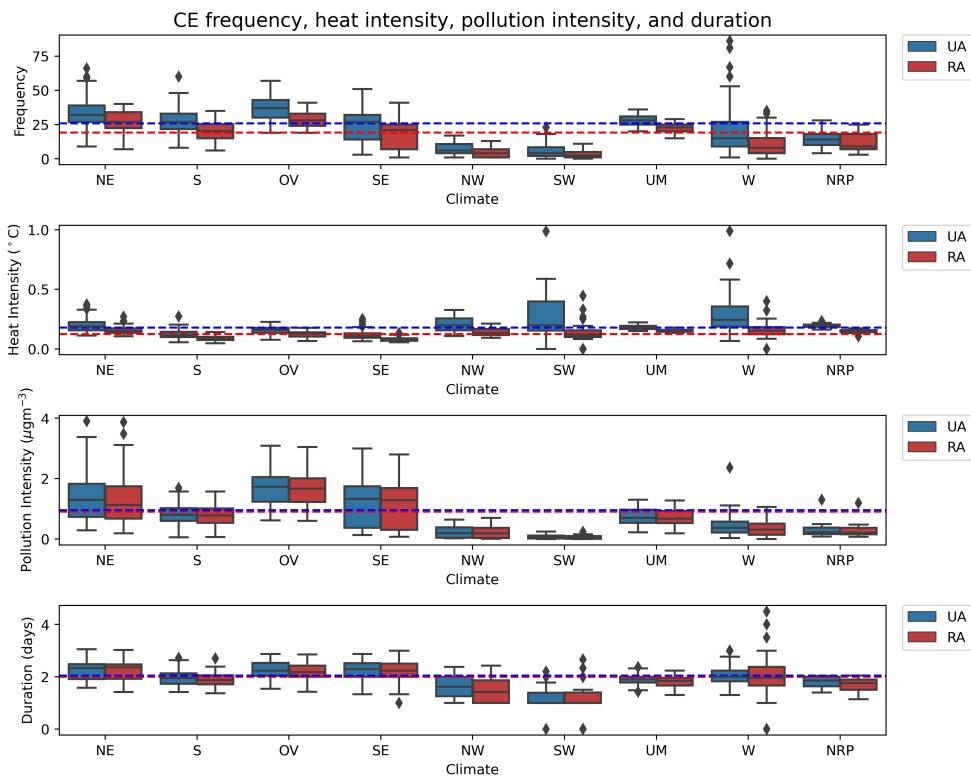

FIG. 6. Regional comparison of CEs frequency, heat intensity, pollution intensity, and duration for UAs and RAs based on NOAA's 9 climate regions: Northeast (NE), South (S), Ohio Valley (OV), Southeast (SE), Northwest (NW), Southwest (SW), Upper Midwest (UM), West (W), Northern Rockies and Plains (NRP). Red dashed line indicates RA average, and blue dashed line indicates UA average. The center line of each box is the median, the box extends from lower to upper quartiles, vertical lines denote 1.5 times the interquartile range, and diamonds are outliers.

Cities where UA frequency was less than RA frequency were Santa Cruz, CA (25% less), Beckley, WV (19.52% less), Grants Pass, OR (12% less), and San Jose, CA (4.35% less). Outliers where UA frequency was greater than RA frequency were Bakersfield, CA (212% greater) and Hilton Head, SC (144% greater). These areas, ex-cept for Hilton Head, are cities primarily surrounded by mountains, and further research could explore the effect of mountainous RAs on UA HWs. There were no significant outliers in comparisons between UA and RA heat intensity and duration.

*b. $PM_{2.5}$ pollution episodes*

The results of AP episode analysis showed that there was some evidence of an "urban pollution island" effect in U.S. cities within AP episode frequency and duration. The variation between regions is due to different sources of pollution, such as $PM_{2.5}$ from wildfires or industrial activities. Further research could evaluate the sources of pollution and how different sources affect AP episodes. Additionally, future research could analyze the trends of AP episodes frequency, intensity, and duration, as local, state, and national legislation controlling pollutants becomes stricter.

It was found that although there were the greatest frequencies and durations of AP episodes in the Northeast, Ohio Valley, and Southeast regions, the differences in AP characteristics between UAs and RAs are small in these regions. This could be a result of non-urban pollution sources, such as industrial activity in both UAs and RAs. This could also be due to circulation patterns, as a study found that a decrease in mixing height and wind speeds in the eastern U.S. resulted in greater concentrations of $PM_{2.5}$ (Dawson et al. 2009).

Significant outliers where UA frequency was greater than RA frequency were Bremerton, WA (650% greater) and Santa Barbara, CA (570% greater). An outlier where UA intensity was greater than RA intensity was San Francisco-Oakland, CA (670% greater), and an outlier where UA duration was greater than RA duration was Bremerton, WA (253% greater). It is notable that these outliers are not in the regions where there were the most frequent, pollution intense, and longest lasting AP episodes, which suggests specific drivers within these cities, which further research could explore. Furthermore, further research could be conducted to understand the effect of wildfire events on pollution levels in these and other western cities.

*c. Compound heat wave and $PM_{2.5}$ pollution episodes*

It was found that CEs were the greatest in frequency in the Northeast and Ohio Valley regions, the greatest in heat intensity in the West and Southwest regions, and the greatest in pollution intensity and duration in the Northeast, Ohio Valley, and Southeast regions. Similarly, the greatest HW frequencies, intensities, and durations were in the West and Southwest regions, the greatest AP episode frequencies and intensities were in the Northeast, Ohio Valley, and Southeast regions, and the longest lasting AP episodes were in the Ohio Valley and Southeast regions. It can be concluded that HWs (AP episodes) had a greater impact on CEs than AP episodes (HWs) in regions with more prominent HWs (AP episodes).

The average heat and pollution intensities of CEs were less than the average heat intensity in HWs and average pollution intensity in AP episodes. This indicates that the most intense and longest lasting HWs are not the cause of the most intense and longest lasting AP episodes. Further research could explore the amount that an HW in a UA can amplify $PM_{2.5}$ levels. This is also consistent with the greatest levels of $PM_{2.5}$ pollution being a result of wildfire events – in some parts of the West, wildfire events can account for up to half of $PM_{2.5}$ levels (Burke et al. 2021).

The most notable outlier from the CE analysis was Sante Fe, NM, where there were zero CEs in either the UA or RA. Outliers where frequency was greater in UA than RA were Seattle, WA (1000% greater) and San Jose, CA (800% greater). These significant differences between UA and RA are a result of several CEs in the UA but one (Seattle) or two (San Jose) in the RA. An outlier where pollution intensity was greater in UA than RA was Bremerton, WA (470% greater). As with HWs and AP episodes, outliers were more common in the West and Northwest regions, and future research could focus on these areas in greater detail.

## 5. Conclusion

The purpose of this study was to provide an understanding of how heat waves (HWs) and $PM_{2.5}$ air pollution (AP) episodes vary in cities across the CONUS and to explore compound HW and AP episodes (CEs), where levels of $PM_{2.5}$ are elevated during an HW. For both urban and rural areas, it was found that HWs were the most frequent, intense, and longest lasting in the West and Southwest regions, AP episodes were the most frequent and longest lasting in the Northeast, Ohio Valley, and Southeast regions, AP episodes were the most intense in the Northern Rockies and Plains and Upper Midwest regions. It was concluded that HWs (AP episodes) had a greater impact on CEs than AP episodes (HWs) in regions with more prominent HWs (AP episodes).

This study motivates further research to better understand the way HWs and AP episodes compound, especially in the urban environment. As large populations in cities are impacted by these events, it is crucial to understand the health impacts of extreme heat and extreme levels of pollution. Future work could examine the trends in the frequency, intensity, and duration of HWs, AP episodes, and CEs, and provide a better understanding of the impacts of background climate on these events in cities.

*Acknowledgments.* We would like to thank Alex Marmo and Dr. Daphne LaDue for organizing the National Weather Center REU program. This material is based upon work supported by the National Science Foundation under Grant No. AGS-2050267.